\documentclass[journal]{IEEEtran}

\ifCLASSINFOpdf
\else
   \usepackage[dvips]{graphicx}
\fi
\usepackage{url}

\def\BibTeX{{\rm B\kern-.05em{\sc i\kern-.025em b}\kern-.08em
    T\kern-.1667em\lower.7ex\hbox{E}\kern-.125emX}}

\usepackage{cite}
\usepackage{amsmath,amssymb,amsfonts}
\usepackage{algorithmic}
\usepackage{graphicx}
\usepackage{textcomp}
\usepackage{caption}
\usepackage{xcolor}
\usepackage{color}
\usepackage{hyperref}
\usepackage{multirow}
\usepackage{subcaption}
\usepackage{fancyhdr}

\begin{document}

\title{Optimizing Multi-Taper Features for Deep Speaker Verification}

\author{Xuechen Liu, Md Sahidullah, \IEEEmembership{Member, IEEE}, and Tomi Kinnunen, \IEEEmembership{Member, IEEE}

\thanks{This project was partially funded by Academy of Finland (project 309629) and Inria Nancy Grand Est.}
\thanks{X. Liu and M. Sahidullah are with Universit\'{e} de Lorraine, CNRS, Inria, LORIA, F-54000, Nancy, France (e-mail: xuechen.liu@inria.fr, md.sahidullah@inria.fr).}
\thanks{X. Liu and T. Kinnunen are with School of Computing, University of Eastern Finland, FI-80101, Joensuu, Finland (e-mail: tomi.kinnunen@uef.fi).}
}

\markboth{Journal of \LaTeX\ Class Files, Vol. 14, No. 8, August 2015}
{Shell \MakeLowercase{\textit{et al.}}: Bare Demo of IEEEtran.cls for IEEE Journals}
\maketitle

\begin{abstract}
Multi-taper estimators provide low-variance power spectrum estimates that can be used in place of the windowed discrete Fourier transform (DFT) to extract speech features such as mel-frequency cepstral coefficients (MFCCs). Even if past work has reported promising automatic speaker verification (ASV) results with Gaussian mixture model-based classifiers, the performance of multi-taper MFCCs with deep ASV systems remains an open question. Instead of a static-taper design, we propose to optimize the multi-taper estimator jointly with a deep neural network trained for ASV tasks. With a maximum improvement on the SITW corpus of 25.8\% in terms of equal error rate over the static-taper, our method helps preserve a balanced level of leakage and variance, providing more robustness.
\end{abstract}

\thispagestyle{fancy}

\fancyhf{}

\renewcommand{\headrulewidth}{0pt}

\chead{\small Accepted to be Published in IEEE Signal Processing Letters}

\pagestyle{empty}

\cfoot{\scriptsize \copyright 2021 IEEE. Personal use of this material is permitted.  Permission from IEEE must be obtained for all other uses, in any current or future media, including reprinting/republishing this material for advertising or promotional purposes, creating new collective works, for resale or redistribution to servers or lists, or reuse of any copyrighted component of this work in other works.}


\IEEEpeerreviewmaketitle

\section{Introduction}
\label{sec:intro}

\IEEEPARstart{F}{eature} extractor is a critical component of speech processing systems. It converts a raw waveform into features that feed task-specific models. In many tasks, including automatic speaker verification (ASV) \cite{asv_2015}, the most widely-used features are the \emph{mel-frequency cepstral coefficients} (MFCCs) computed from a short-term spectral representation—usually, the windowed \emph{discrete Fourier transform} (DFT) \cite{Harris1978-windowed-DFT}.

While MFCCs perform well under matched conditions, they lack robustness to data variations. Various methods are available to improve their robustness, such as feature normalization \cite{Huang_spokenlang2001,rasta1994} applied \emph{after} MFCC extraction. In addition, the MFCC extractor itself can be improved. In \cite{multitaper2012}, a spectral estimator based on multiple windows was used in place of the single-window DFT. Such \emph{multi-taper} spectrum estimator \cite{thomson} addresses a specific shortcoming --- high variance. Here, a single frame of speech is viewed as a realization of a stationary stochastic process, and `variance' refers to the variation in power-spectral estimates. Given the ubiquitous role of the power spectrum in different speech front-ends, multi-tapers can also be used to enhance other features, such as \emph{perceptual linear predictive} (PLP) features \cite{multitaper_mfcc_plp2013, multitaper_mfcc_plp2012}.

A multi-taper power spectrum estimator is simply a weighted average of many windowed DFT power spectrum estimates (\emph{sub-spectra}) obtained with carefully designed window functions (\emph{tapers}) and their associated weights. While there are different approaches to design optimal multi-tapers \cite{swce2009,thomson}, they typically rely upon assumptions of the stochastic process, rather than the downstream application. In ASV, for instance, we do not extract identity cues from a \emph{single} speech frame --- the domain of the short-term power spectrum --- but multiple frames, i.e. an utterance. Thus, while application-independent multi-taper design is the standard one, it is plausible that the assumed stochastic process properties are not compatible with the given downstream task or classifier.

Experiments with \emph{Gaussian mixture model} (GMM) based classifiers in \cite{multitaper_mfcc_plp2013, multitaper_mfcc_plp2012, multitaper_mfcc_plp2014} indicate that the multi-taper spectrum estimator is a simple yet effective method to improve ASV accuracy. Nonetheless, the community has recently shifted its focus to \emph{deep neural networks} (DNNs; for a comprehensive survey, refer to \cite{Bai2021-asv-dnn-review}). This raises a question whether the earlier positive findings can be generalized to modern ASV models, which motivated the present work. 

One crucial difference between GMMs and DNNs lies in their ability to model larger temporal contexts. GMMs cannot handle high-dimensional data, as it would require additional dimensionality reduction, diagonal covariances, or limiting the number of frames that can be used for feature extraction. Many DNN architectures (e.g. models with recurrent, dilated convolutional or time-delay layers), however, can cope with an extended temporal context without issues. A successful example is the \emph{time-delay neural network} (TDNN) architecture \cite{2015_tdnn} used in \emph{x-vector} model \cite{xvector2018}, whose variations and extensions \cite{Snyder_etdnn_2019, ecapa-tdnn} currently form the standard ASV baselines. 

The spectrum variance reduction provided by traditional multi-tapers on individual frames might not perfectly combine with a TDNN. We hypothesize that better spectral features could be obtained by optimizing the multi-taper estimator \emph{for an ASV task directly}. While in GMM-based approaches `features' and `classifiers' are often treated separately, DNNs enable their joint optimization. Although this is the overall sentiment in \textit{end-to-end} learning \cite{dvector_rawwave_cnn2017, e2e_waveform_cnn2018, e2e_waveform_cnn2018_2, sincnet2018}, our starting points are DFT-based spectral representation and MFCCs rather than the raw waveform. Therefore, our feature extractor design retains the familiar processing elements and enables one-to-one comparisons with conventional MFCCs obtained either from a single-window DFT, or hand-crafted static multi-taper spectrum estimators. Besides quantitative ASV evaluation on three different datasets, we investigate the spectral and statistical properties of the learned multi-taper estimator.

\section{Optimized Multi-taper Spectral Estimator}
\label{sec:guidelines}

\subsection{Multi-taper}
Let $\boldsymbol{x} = [x(0), \dots, x(N-1)]$ to denote a short speech frame of $N$ samples. The windowed DFT \cite{Harris1978-windowed-DFT} is defined by
\begin{equation}
    \hat{S}(f) = \Bigg|\sum_{t=0}^{N-1}w(t)x(t)e^{-i2{\pi}tf/N}\Bigg|^{2},
\label{eq:windowdft}
\end{equation}
where $\hat{S}(f)$ denotes the estimated power spectrum and $f=0,\dots,N-1$ is the DFT frequency bin. Additionally, $w(t)$ is the window function (taper) --- in this work, the \textit{Hamming} window with $w(t) = 0.54-0.46\cos(2{\pi}t/N)$ for $0 \leq t<N$ (and $w(t)=0$ for other $t$). The primary purpose of the window is to reduce \emph{spectral leakage} compared to the rectangular window, also known as `no windowing'. Nonetheless, the variance of the spectrum estimates provided by \eqref{eq:windowdft} remains high. The multi-taper spectral estimator \cite{thomson} aims at reducing the variance through multiple, weighted DFT power spectrum estimates:
\begin{equation}
    \hat{S}(f) = \sum_{j=1}^{K}\lambda(j)\Bigg|\sum_{t=0}^{N-1}w_{j}(t)x(t)e^{-i2{\pi}tf/N}\Bigg|^{2},
\label{eq:multi-taper}
\end{equation}
where $K$ is the number of tapers, $j=1,\dots,K$ denotes the taper index and  $\boldsymbol{w}_{j} = [w_{j}(0),\dots, w_{j}(N-1)]$ represents $j^\text{th}$ taper, with its associated weight $\lambda(j) > 0$. Windowed DFT in \eqref{eq:windowdft} is obtained as a special case with $K=1$, $\lambda=1$ and $\boldsymbol{w}_1$ set as the Hamming window.

Averaging reduces variance by making the resulting spectrum less susceptible to small within-frame data variation. After a suitable set of tapers has been selected, their number ($K$) can be selected to trade-off between two conflicting criteria of variance reduction (high $K$) and bias reduction (low $K$). A high value of $K$ implies a `rigid' spectrum that smears detail but provides a (statistically) stable representation; a low $K$, in turn, retains more detail but is susceptible to perturbations.
The choice of $K$ typically requires some experimentation in a given downstream task.

In previous work, the tapers and their weights were set in a hand-crafted manner. A detailed study in GMM-based speaker recognition was conducted in \cite{multitaper2012}, where three different types of taper designs were considered. 
Based on preliminary ASV experiments, in this work, we focus on \emph{sine-weighted cepstral estimator} (SWCE) \cite{swce2009}, where both the tapers and their weights are provided by closed-form equations:

\begin{align}
\label{eq:swce-eqs}
    \begin{aligned}
        w_{j}(t) & =   \sqrt{2/(N+1)}\sin\big[2{\pi}jt/(N+1)\big] \\     \lambda(j) & =\sin\big[2{\pi}j/(N+1)\big]/\sum_{k=0}^{K}\sin\big[2{\pi}k/(N+1))\big].      
    \end{aligned}
\end{align}

Readers may refer to \cite{thomson} for further details on multi-tapers, which are out of our scope. In general, the tapers and their weights are designed to uncorrelate the estimation errors between the sub-spectra. As noted above, however, the tapers are designed for short-term stationary signals (frames). Hence, such design ignores both the temporal context and interaction with the downstream model, which is DNN-based ASV here.

\subsection{Data-driven multi-taper}

As explained above, we attempted to learn the multi-taper spectrum estimator jointly with a downstream system --- specifically, a DNN-based speaker embedding extractor.
To this end, we treated the taper weights $\boldsymbol{\lambda} = \lambda(j), j=1,...,K$ as part of the neural network parameters. They were updated jointly with the TDNN parameters, denoted by $\boldsymbol{W}$. As an example, using first-order gradient descent, the model parameters are updated by, 

\begin{align}
\label{eq:sgd-swce}
\begin{aligned}
    \boldsymbol{W}_{nk} &\leftarrow \boldsymbol{W}_{nk} - \eta_{n} * \frac{\partial J_\text{loss}}{\partial \boldsymbol{W}_{nk}} \\
    \boldsymbol{\lambda}_{nk} &\leftarrow \boldsymbol{\lambda}_{nk} - \eta_{n} * \frac{\partial J_\text{loss}}{\partial \boldsymbol{\lambda}_{nk}},
\end{aligned}
\end{align}
where $J_\text{loss}$ is the objective function of the neural network (here, additive angular softmax \cite{angular_softmax2019} between the output of the network and speaker labels), and $\eta$ denotes the learning rate; $n$ and $k$ are epoch and iteration indices, respectively. The gradient with respect to network and taper weights is computed based on chain rule. In this study, we used Adam optimizer~\cite{adam}.

We optimized only for the taper \emph{weights}, while the tapers $\boldsymbol{w}_{j}$ remained static, which allowed the weights to be treated as `leaf' scalar nodes in the computational graph during optimization, making optimization efficient without introducing additional training parameters and neural layers. Moreover, it avoided introducing an excessive number of additional taper parameters ($K \cdot N$, where $N$ is the number of spectral bins), which could have made the joint learning challenging in terms of finding an optimal solution. Our preliminary ASV experiments with learnable taper functions (represented by the vectors $\boldsymbol{w}_{j}$) indicated less promising result; hence, this direction was not pursued further.

In addition to the choice of the number of tapers $K$, other important design considerations include weight initialization and their non-negativity constraints. Following \cite{learnable_mfcc2021}, we considered two initialization approaches. The first used random weights from a standard normal distribution, and the latter initialized the weights using \eqref{eq:swce-eqs}.
Note that since \eqref{eq:multi-taper} is a power spectrum estimator, we require $\lambda(j) > 0$ as a constraint. Inspired by previous works on different types of tapers \cite{multitaper_mfcc_plp2012}, we additionally constrained the sum of weights to unity ($\sum_{j=1}^K \lambda(j) = 1$). To this end, the updated weights are processed with rectified linear unit (ReLU) \cite{relu2010} activation function to enforce positivity, followed by length normalization $\boldsymbol{\lambda} \leftarrow \boldsymbol{\lambda}/\Vert \boldsymbol{\lambda}\Vert_1$ to enforce the latter constraint. The choice of $\ell_1$ norm is inspired in part by sparsity considerations. We refer to such constraint by simply noting the ReLU function in the following sections.

\begin{table*}[htbp]
  \renewcommand{\arraystretch}{1.1}
  \centering
  \caption{Speaker verification results on different evaluation sets.}
  \begin{tabular}{|c|c|ccc|cc|cc|cc|}
    \hline
    & & & & & \multicolumn{2}{|c|}{Voxceleb1-test} & \multicolumn{2}{c|}{SITW-EVAL} & \multicolumn{2}{c|}{ASVSpoof2019-LA} \\ \hline
    Type & Taper & Num. tapers & Weight init. & Weight constraint & EER(\%) & MinDCF & EER(\%) & MinDCF & EER(\%) & MinDCF \\ \hline
    \multirow{2}{2.75em}{Static} & DFT & - & - & - & 2.01 & 0.2344 & 2.93 & 0.2901 & 1.52 & 0.149 \\
    & SWCE & 8 & - & - & 2.12 & 0.2663 & 3.72 & 0.3214 & 1.32 & 0.1546 \\ \hline
    \multirow{8}{2.75em}{Data-driven} & \multirow{8}{2.75em}{SWCE} & 8 & Gaussian & None & 2.23 & 0.2364 & 3.19 & 0.3321 & 1.60 & \textbf{0.1405} \\ 
    & & 8 & Gaussian & \textit{ReLU} & \textbf{1.96} & 0.2473 & 3.11 & 0.2939 & 1.38 & 0.1500 \\
    & & 20 & Gaussian & \textit{ReLU} & 2.01 & 0.2503 & 2.95 & 0.2864 & 1.34 & 0.2403 \\ \cline{3-11}
    & & 8 & SWCE & None & \textbf{1.96} & \textbf{0.2209} & \textbf{2.78} & \textbf{0.2862} & \textbf{1.20} & 0.1570 \\
    & & 8 & SWCE & \textit{ReLU} & 2.12 & 0.2596 & 2.87 & 0.2932 & 1.50 & 0.1485 \\
    & & 20 & SWCE & \textit{ReLU} & 2.33 & 0.3213 & 3.74 & 0.3583 & 1.45 & 0.1561 \\
    \cline{3-11}
    & & 2 & Gaussian & None & 1.98 & 0.2497 & 3.21 & 0.3003 & 1.34 & 0.1325 \\
    & & 2 & SWCE & None & \textbf{1.95} & 0.2559 & 3.06 & 0.2969 & 1.42 & 0.1377 \\
    \hline
  \end{tabular}
\label{tab:results}
\end{table*}

\begin{table}[h]
  \renewcommand{\arraystretch}{1.0}
  \centering
  \caption{Spectral analysis statistics on input synthetic signal for different multi-taper estimators. The number of tapers for systems covered in this table is 8.}
  \begin{tabular}{|c|c|c|cc|}
    \hline
    Type & Taper & Weight init./constraint & Distance. & Width(Hz) \\ \hline
    \multirow{2}{2.7em}{Static} & DFT & - & 0.21 & 39.26 \\
    & SWCE & - & 1.04 & 314.15 \\ \hline
    \multirow{4}{2.7em}{Data-driven} & \multirow{4}{2.7em}{SWCE} & Gaussian/None & 0.31 & 196.35 \\ 
    & & Gaussian/\textit{ReLU} & 0.29 & 196.35 \\ \cline{3-5}
    & & SWCE/None & 0.36 & 196.35 \\ 
    & & SWCE/\textit{ReLU} & 0.34 & 235.62 \\ \hline
  \end{tabular}
\label{tab:analysis}
\end{table}

\section{Experimental Setup}
\label{sec:experiments}

\textbf{Data}. We reported ASV performance on three different datasets. The first one was \emph{Voxceleb1-test}, a 40-speaker \emph{test} partition from Voxceleb1 following \cite{Nagrani_vox1_2017}. We also included \emph{eval} partition of \emph{speakers in the wild} (SITW) corpus \cite{Mclaren_sitw2016}, under core-core condition (\emph{SITW-EVAL}) and logical access (LA) scenario of ASVspoof 2019 \cite{asvspoof2019} with only bonafide imposters from its ASV protocol (\emph{ASVSpoof2019-LA}). The three datasets have diverse qualities; \emph{ASVSpoof2019-LA} has the lowest acoustic mismatch between the enrollment and the test data due to its relatively clean and highly-controlled recording conditions. The other two datasets are more similar to one another, as both contain audio extracted from videos.

The speaker embedding extractors were trained on the VoxCeleb. We used the \textit{dev} partitions~\cite{Nagrani_vox1_2017, Chung_vox2_2018} consisting of 7205 speakers after removing speakers overlapped with SITW.

\textbf{Feature extractors}. MFCCs obtained with conventional Hamming window-based DFT and with SWCE-based multi-taper spectrum estimator \cite{swce2009} formed our baselines. For the proposed data-driven multi-taper MFCCs, we contrasted the two initialization methods explained above. 
We also addressed the impact of the proposed ReLU update rule and noted the results for the selected number of tapers. In all cases, the acoustic features that feed the neural network were 40 MFCCs computed with the same number of the mel filters.

\textbf{Speaker embedding extractor}. \emph{Extended x-vector} based on \emph{time-delayed neural network} (E-TDNN) \cite{Snyder_etdnn_2019} served as the speaker embedding extractor, which showed promising performance over the basic x-vector model \cite{xvector2018}. In addition, we replaced the conventional statistics pooling in the original network with attentive statistics pooling \cite{astats_pooling} and trained the network using the \emph{additive angular margin softmax} loss function \cite{angular_softmax2019}. We extracted a 512-dimensional speaker embedding from the first fully-connected layer after statistics pooling for each input utterance.

\textbf{Classifier back-end}. For each corpus, we trained a probabilistic linear discriminant (PLDA) back-end classifier using speaker embeddings from the trained speaker embedding extractor. The extracted embeddings were length-normalized and centered using a 200-dimensional linear discriminant analyzer (LDA) prior to the PLDA estimation. The subspace size of PLDA was the same as that of LDA.

\textbf{Metrics}. We evaluated the speaker verification performance with \emph{equal error rate} (EER) and minimum detection cost function (minDCF). Target speaker prior for minDCF was $p_\text{tar} = 0.01$. \textit{Detection error trade-off} (DET) curves on SITW-EVAL were also provided. We used Kaldi \cite{kaldi} to compute EERs and minDCFs and BOSARIS \cite{bosaris} to draw DET plots.

\begin{figure}[t]
\centering
\centerline{\includegraphics[width=7cm]{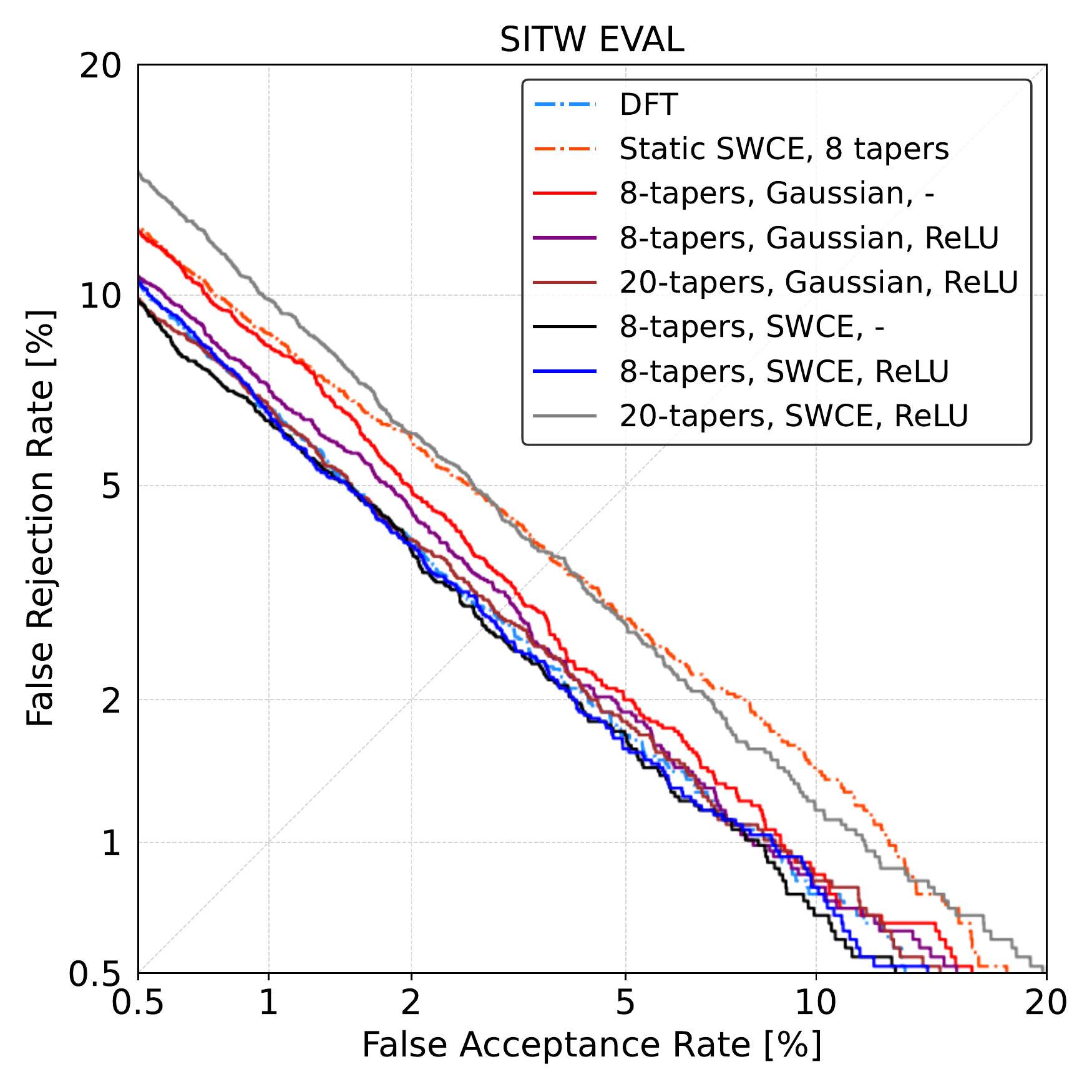}}
\hfill
\caption{DET curves on SITW-EVAL. Proposed spectral estimators are marked with (num. tapers, weight init., weight constraint) from Table \ref{tab:results}.}
\label{fig:det}
\vspace{-0.25cm}
\end{figure}

\section{Experimental Results}
\label{sec:results}

\subsection{Speaker Verification Results}
Table \ref{tab:results} shows the results for the baseline MFCCs, static multi-taper MFCCs, and the proposed data-driven variants. Similar to our previous findings in \cite{Xuechen_feature2020} using a smaller-scale training set, static SWCE with eight tapers did not outperform baseline MFCCs on the \emph{Voxceleb1-test}. However, it yielded slightly lower EER on \emph{ASVSpoof2019-LA} than the baseline, which indicated its potentiality in test conditions with a lower mismatch between enrollment and test.

For condition, namely \emph{Voxceleb1-test}, data-driven multi-taper systems did not show a substantial advantage over the Hamming window but demonstrated consistent improvement over static multi-taper in most cases. SWCE weight initialization yielded the lowest EERs and minDCFs. For \emph{ASVSpoof2019-LA}, best system outperformed static SWCE by relatively 9.1\% on EER and 5.7\% on minDCF. The data-driven taper exhibited a noticeable performance gain on \emph{SITW-EVAL} with a 25.8\% relative EER improvement compared to static SWCE, which indicate that data-driven approaches can improve robustness. 

Adding ReLU generally improved the ASV performance of systems with Gaussian weight initialization, which was not the case for SWCE kernel initialization, where such addition degraded performance in all categories except for minDCF for \textit{ASVSpoof2019-LA}. Our proposed updating approach might over-fit the weights since kernel initialization already sets a proper starting point for learning. This can be observed for both $K=8$ and $K=20$. Moreover, we found from pilot experiments that using larger number of tapers (e.g. 32) degraded the performance\footnote{EER and minDCF are 3.56\%/0.4091 for Gaussian/\textit{ReLU} and 2.82\%/0.3443 for SWCE/\textit{ReLU} on VoxCeleb1-test for $K=32$.}, which may due to over-smoothed spectrum. With increased computational time, we limited our experiments to maximum of $K=20$ tapers.

The above findings are in line with the DET curve (Fig.~\ref{fig:det}) in most operating points. The advantage of learned 8-taper systems compared to static ones is apparent in regions with less constraint on false alarms, especially with kernel initialization.

\begin{figure}[t]
    \centering
    \includegraphics[width=0.5\textwidth]{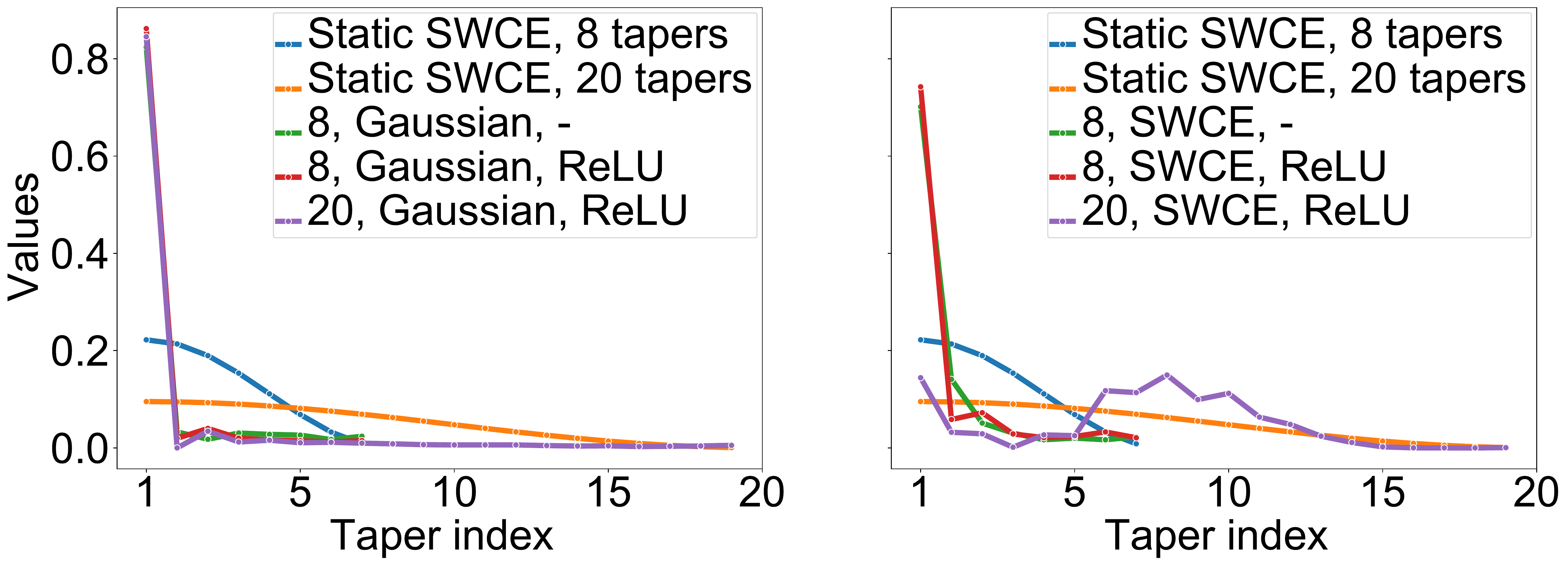}
    \caption{Weights of different static and data-driven multi-taper estimators. Proposed spectral estimators are marked with (num. tapers, weight init., weight constraint) from Table \ref{tab:results}.}
    \label{fig:weights}
\vspace{-0.5cm}
\end{figure}

\subsection{Analysis}

Two analytic studies appeared to be of interest: 1) a study on learned taper weight values, compared to hand-crafted settings; 2) a study on spectral leakage. The former can give an alternative design choice for multi-taper estimators, while the latter can bridge the statistical importance of the estimators and deep ASV performance. 

\textbf{Learned taper weights}. Figure \ref{fig:weights} shows the weight values of different learned spectral estimators, including the baseline static SWCE. Among all 8-taper learned multi-taper estimators, weight values are heavily concentrated on the first two tapers, with the remaining weights being close to zero. In light of the DNN-based ASV results, a lower number of tapers may be better for robustness. To further validate this hypothesis, we conducted two additional experiments with two tapers only with Gaussian and SWCE initialization, respectively. The results of those two experiments are shown at the end of Table \ref{tab:results}. They show that using such a low number of tapers does not exacerbate ASV performance. In fact, the one with SWCE weight initialization reached the lowest EER on \textit{Voxceleb1-test}, outperforming static SWCE by 8.0\%.

\begin{figure}[t]
  \begin{subfigure}[b]{\linewidth}
    \centering
    \includegraphics[width=0.85\linewidth]{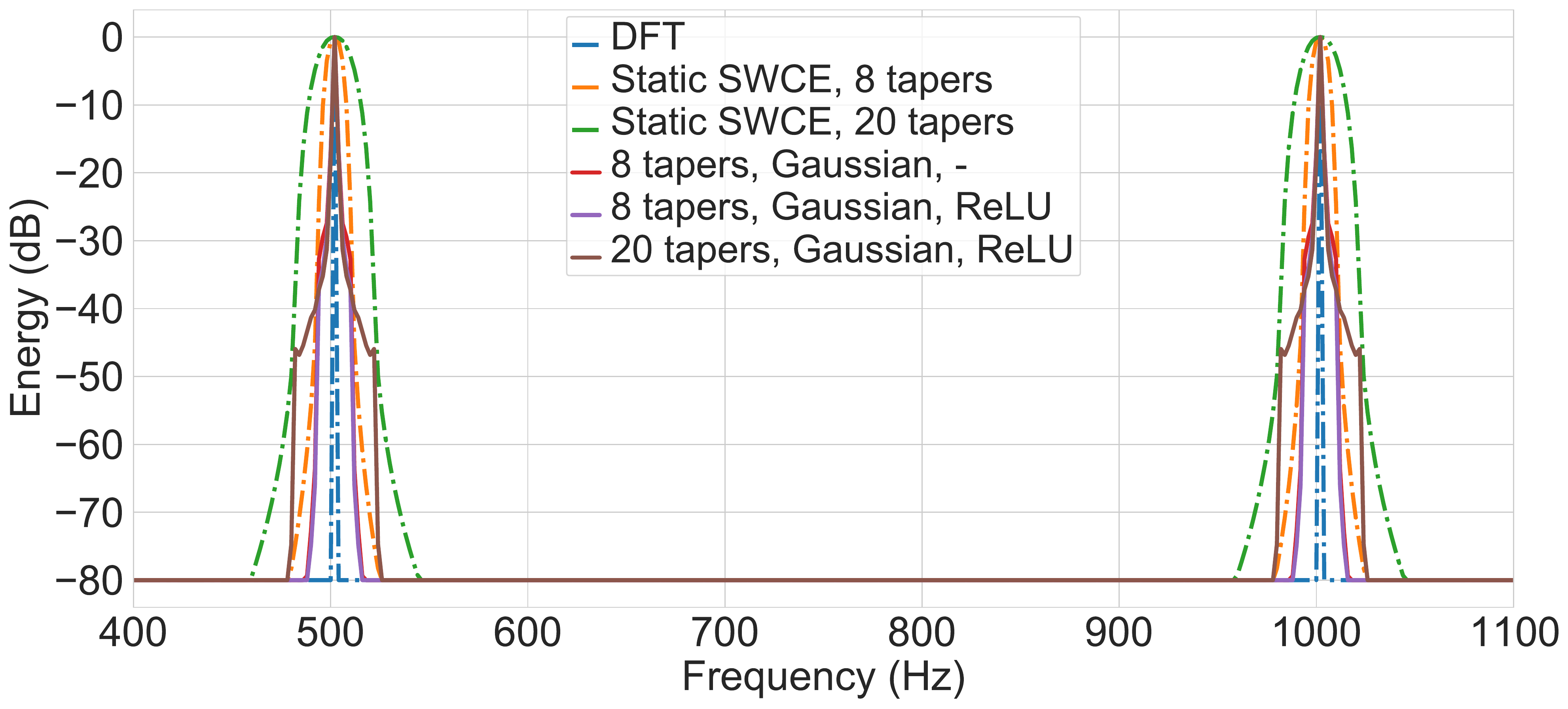} 
    \label{fig7:random} 
  \end{subfigure} 
  \vfill
  \begin{subfigure}[b]{\linewidth}
    \centering
    \includegraphics[width=0.85\linewidth]{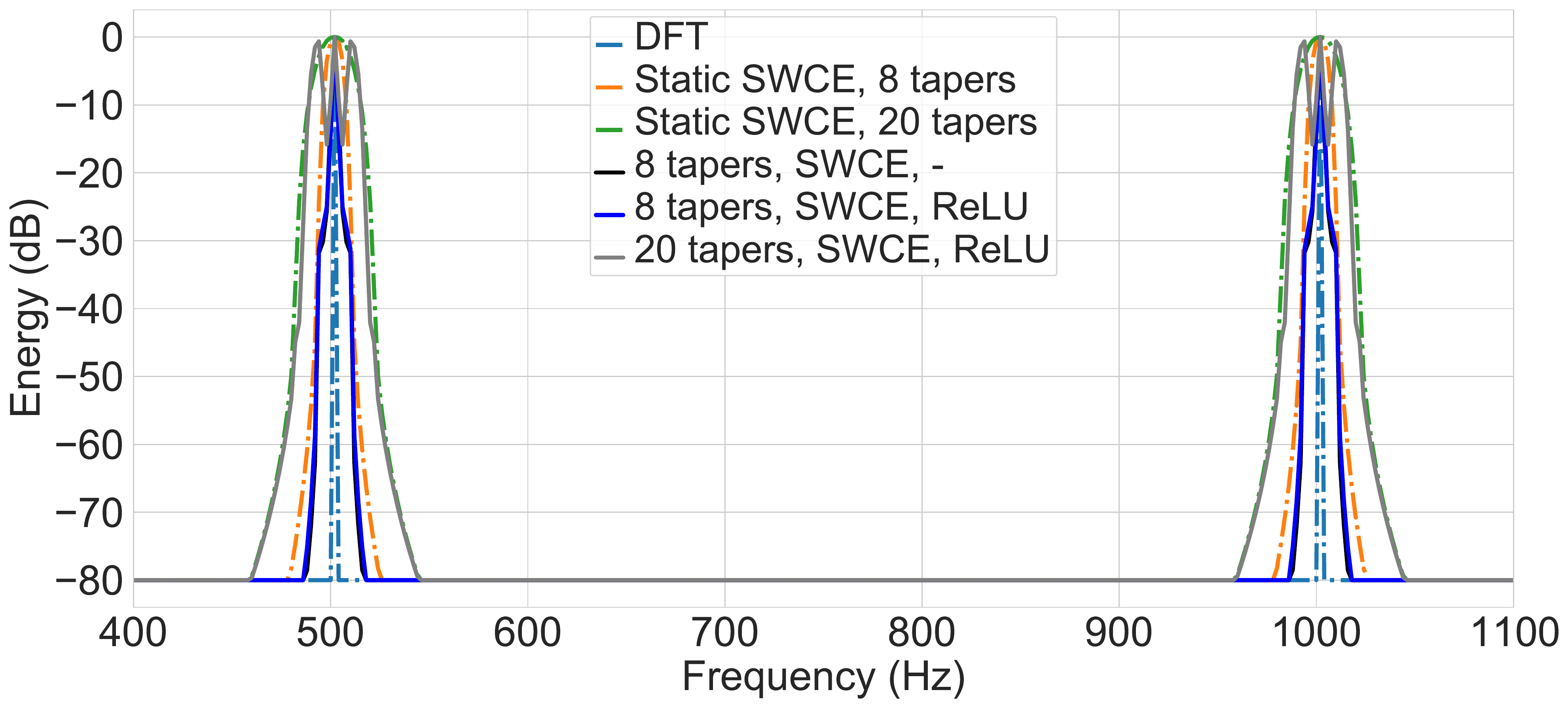} 
    \label{fig7:swce} 
  \end{subfigure} 
\caption{Spectral representation from sinusoids. Proposed spectral estimators are marked with (num. tapers, weight init., weight constraint) from Table \ref{tab:results}.}
\label{fig:spectra}
\vspace{-0.5cm}
\end{figure}

\textbf{Spectral leakage}. To measure relative leakage experimentally, we generated synthetic signal, by $s_{n}(t) = \sin(2\pi nf_{s}t / N_{\mathrm{FFT}})$, where $f_{s}$ is sampling rate (16~kHz here), $n$ is the frequency bin index that made to unit amplitude of spectral energy, $\sin$ denotes sinusoidal function and $N_{\mathrm{FFT}}$ is the number of FFT bins (512 here). The final signal is the sum of $s_{n}(t)$ of different frequencies. We measured the performance of different estimators by two metrics. The first one was the spectral difference from ground truth measured through the Itakura-Saito distance \cite{is_distance}. Second, we defined and measured the attenuation width where the spectral energies were sufficiently low. This is expressed as $w = (n_{\mathrm{right}} - n_{\mathrm{left}}) * f_{s} / N_{\mathrm{FFT}}$ in Hz, where $n_{\mathrm{right}}$ and $n_{\mathrm{left}}$ are edge points where the spectral power is 80~dB below unity gain. We chose 500~Hz and 1~kHz as center frequencies, referring to the average values of the first two formant frequencies \cite{phonetics_textbook}. Spectra returned by all estimators including two baselines are visualized in Fig. \ref{fig:spectra} and Table \ref{tab:analysis} shows the leakage statistics of the static and the 8-taper systems with respect to ground truth. The leakage indicated in the figure shows that the better-performed data-driven systems return a certain level of leakage that lies between their static counterparts and ground truth. Numbers returned by best-performing systems lies in between static SWCE and ground truth representation, indicating that while lower spectral leakage is expected, a trade-off between certain levels of leakage and perturbation must occur in order to reach better ASV performance.

\section{Conclusion}
We re-evaluated static multi-taper spectral estimator for speaker verification with DNN-based speaker embedding extractor and proposed optimization schemes, enabling joint learning of taper weight values with the DNN. We then investigated the effect of kernel initialization using the static counterparts. The proposed optimized multi-taper features show promising speaker verification performance and a high level of robustness on varieties of speech corpora. Further analysis shows that the learned multi-taper implicitly maintains a decent trade-off between spectral leakage and variance, corresponding to an improved ASV performance.

\bibliographystyle{IEEEbib}
\bibliography{strings,refs}

\begin{thebibliography}{10}

\bibitem{asv_2015}
J.~H.L. Hansen and T.~Hasan,
\newblock ``Speaker recognition by machines and humans: A tutorial review,''
\newblock {\em IEEE Signal Processing Magazine}, vol. 32, no. 6, pp. 74--99,
  2015.

\bibitem{Harris1978-windowed-DFT}
F.J. Harris,
\newblock ``On the use of windows for harmonic analysis with the discrete
  {Fourier} transform,''
\newblock {\em Proceedings of the IEEE}, vol. 66, no. 1, pp. 51--83, 1978.

\bibitem{Huang_spokenlang2001}
X.~Huang, A.~Acero, H.~Hon, and R.~Reddy,
\newblock {\em Spoken Language Processing: A Guide to Theory, Algorithm, and
  System Development},
\newblock Prentice Hall PTR, USA, 1st edition, 2001.

\bibitem{rasta1994}
H.~{Hermansky} and N.~{Morgan},
\newblock ``{RASTA} processing of speech,''
\newblock {\em IEEE Transactions on Speech and Audio Processing}, vol. 2, no.
  4, pp. 578--589, 1994.

\bibitem{multitaper2012}
T.~{Kinnunen}, R.~{Saeidi}, F.~{Sedlak}, K.~A. {Lee}, J.~{Sandberg},
  M.~{Hansson-Sandsten}, and H.~{Li},
\newblock ``Low-variance multitaper {MFCC} features: A case study in robust
  speaker verification,''
\newblock {\em IEEE Transactions on Audio, Speech, and Language Processing},
  vol. 20, no. 7, pp. 1990--2001, 2012.

\bibitem{thomson}
D.J. Thomson,
\newblock ``Spectrum estimation and harmonic analysis,''
\newblock {\em Proceedings of the IEEE}, vol. 70, no. 9, pp. 1055--1096, 1982.

\bibitem{multitaper_mfcc_plp2013}
M.~J. Alam, T.~Kinnunen, P.~Kenny, P.~Ouellet, and D.~O’Shaughnessy,
\newblock ``Multitaper {MFCC} and {PLP} features for speaker verification using
  i-vectors,''
\newblock {\em Speech Communication}, vol. 55, no. 2, pp. 237--251, 2013.

\bibitem{multitaper_mfcc_plp2012}
M.~Alam, P.~Kenny, and D.~O’Shaughnessy,
\newblock ``Low-variance multitaper mel-frequency cepstral coefficient features
  for speech and speaker recognition systems,''
\newblock {\em Cognitive Computation}, vol. 5, 12 2013.

\bibitem{swce2009}
M.~Hansson-Sandsten and J.~Sandberg,
\newblock ``Optimal cepstrum estimation using multiple windows,''
\newblock in {\em Proc. ICASSP}, 2009, pp. 3077--3080.

\bibitem{multitaper_mfcc_plp2014}
M.~J. Alam, P.~Kenny, P.~Dumouchel, and D.~O'Shaughnessy,
\newblock ``Robust feature extractors for continuous speech recognition,''
\newblock in {\em 2014 22nd European Signal Processing Conference (EUSIPCO)},
  2014, pp. 944--948.

\bibitem{Bai2021-asv-dnn-review}
Z.~Bai and X.~Zhang,
\newblock ``Speaker recognition based on deep learning: An overview,''
\newblock {\em Neural Networks}, vol. 140, pp. 65--99, 2021.

\bibitem{2015_tdnn}
V.~Peddinti, D.~Povey, and S.~Khudanpur,
\newblock ``A time delay neural network architecture for efficient modeling of
  long temporal contexts,''
\newblock in {\em Proc. INTERSPEECH}, 2015.

\bibitem{xvector2018}
D.~Snyder, D.~Garcia-Romero, G.~Sell, D.~Povey, and S.~Khudanpur,
\newblock ``X-vectors: Robust {DNN} embeddings for speaker recognition,''
\newblock in {\em Proc. ICASSP}, 2018, pp. 5329--5333.

\bibitem{Snyder_etdnn_2019}
D.~{Snyder}, D.~{Garcia-Romero}, G.~{Sell}, A.~{McCree}, D.~{Povey}, and
  S.~{Khudanpur},
\newblock ``Speaker recognition for multi-speaker conversations using
  x-vectors,''
\newblock in {\em Proc. ICASSP}, 2019, pp. 5796--5800.

\bibitem{ecapa-tdnn}
B.~Desplanques, J.~Thienpondt, and K.~Demuynck,
\newblock ``{ECAPA-TDNN: Emphasized Channel Attention, Propagation and
  Aggregation in TDNN Based Speaker Verification},''
\newblock in {\em Proc. INTERSPEECH}, 2020, pp. 3830--3834.

\bibitem{dvector_rawwave_cnn2017}
J.~Jung, H.~Heo, I.~Yang, S.~Yoon, H.~Shim, and H.~Yu,
\newblock ``D-vector based speaker verification system using raw waveform
  {CNN},''
\newblock in {\em Proceedings of the 2017 International Seminar on Artificial
  Intelligence, Networking and Information Technology (ANIT 2017)}. 2017/12,
  pp. 126--131, Atlantis Press.

\bibitem{e2e_waveform_cnn2018}
J.~Jung, H.~Heo, I.~Yang, H.~Shim, and H.~Yu,
\newblock ``A complete end-to-end speaker verification system using deep neural
  networks: From raw signals to verification result,''
\newblock in {\em Proc. ICASSP}, 2018, pp. 5349--5353.

\bibitem{e2e_waveform_cnn2018_2}
H.~Muckenhirn, M.~Magimai-Doss, and S.~Marcell,
\newblock ``Towards directly modeling raw speech signal for speaker
  verification using {CNN}s,''
\newblock in {\em Proc. ICASSP}, 2018, pp. 4884--4888.

\bibitem{sincnet2018}
M.~Ravanelli and Y.~Bengio,
\newblock ``Speaker recognition from raw waveform with sincnet,''
\newblock in {\em 2018 IEEE Spoken Language Technology Workshop (SLT)}, 2018,
  pp. 1021--1028.

\bibitem{angular_softmax2019}
J.~Deng, J.~Guo, N.~Xue, and S.~Zafeiriou,
\newblock ``Arcface: Additive angular margin loss for deep face recognition,''
\newblock in {\em 2019 IEEE/CVF Conference on Computer Vision and Pattern
  Recognition (CVPR)}, 2019, pp. 4685--4694.

\bibitem{adam}
D.~P. Kingma and J.~Ba,
\newblock ``Adam: A method for stochastic optimization,''
\newblock in {\em ICLR (Poster)}, 2015.

\bibitem{learnable_mfcc2021}
X.~Liu, M.~Sahidullah, and T.~Kinnunen,
\newblock ``Learnable {MFCC}s for speaker verification,''
\newblock in {\em 2021 IEEE International Symposium on Circuits and Systems
  (ISCAS)}, 2021, pp. 1--5.

\bibitem{relu2010}
V.~Nair and G.~E. Hinton,
\newblock ``Rectified linear units improve restricted boltzmann machines,''
\newblock in {\em Proceedings of the 27th International Conference on
  International Conference on Machine Learning}, Madison, WI, USA, 2010,
  ICML'10, p. 807–814, Omnipress.

\bibitem{Nagrani_vox1_2017}
A.~Nagrani, J.~Chung, and A.~Zisserman,
\newblock ``{VoxCeleb}: A large-scale speaker identification dataset,''
\newblock in {\em Proc. INTERSPEECH}, 2017, pp. 2616--2620.

\bibitem{Mclaren_sitw2016}
M.~McLaren, L.~Ferrer, Diego Castán~L., and A.~Lawson,
\newblock ``The speakers in the wild ({SITW}) speaker recognition database,''
\newblock in {\em Proc. INTERSPEECH}, 2016, pp. 818--822.

\bibitem{asvspoof2019}
A.~{Nautsch}, X.~{Wang}, N.~{Evans}, T.~H. {Kinnunen}, V.~{Vestman},
  M.~{Todisco}, H.~{Delgado}, M.~{Sahidullah}, J.~{Yamagishi}, and K.~A. {Lee},
\newblock ``{ASVspoof} 2019: Spoofing countermeasures for the detection of
  synthesized, converted and replayed speech,''
\newblock {\em IEEE Transactions on Biometrics, Behavior, and Identity
  Science}, vol. 3, no. 2, pp. 252--265, 2021.

\bibitem{Chung_vox2_2018}
J.~S. Chung, A.~Nagrani, and A.~Zisserman,
\newblock ``{VoxCeleb2}: Deep speaker recognition,''
\newblock in {\em Proc. INTERSPEECH}, 2018.

\bibitem{astats_pooling}
K.~Okabe, T.~Koshinaka, and K.~Shinoda,
\newblock ``Attentive statistics pooling for deep speaker embedding,''
\newblock in {\em Proc. INTERSPEECH}, 2018, pp. 2252--2256.

\bibitem{kaldi}
D.~Povey et~al,
\newblock ``The kaldi speech recognition toolkit,''
\newblock in {\em Proc. ASRU}. 2011, IEEE Signal Processing Society.

\bibitem{bosaris}
N.~Br{\"u}mmer and E.~Villiers,
\newblock ``The {BOSARIS} toolkit: Theory, algorithms and code for surviving
  the new {DCF},''
\newblock {\em ArXiv}, vol. abs/1304.2865, 2013.

\bibitem{Xuechen_feature2020}
X.~Liu, M.~Sahidullah, and T.~Kinnunen,
\newblock ``A comparative re-assessment of feature extractors for deep speaker
  embeddings,''
\newblock in {\em Proc. INTERSPEECH}, 2020, pp. 3221--3225.

\bibitem{is_distance}
F~Itakura and S~Saito,
\newblock ``Analysis synthesis telephony based on the maximum likelihood
  method,''
\newblock in {\em Proc. 6th of the International Congress on Acoustics}, 1968,
  pp. C17--C20.

\bibitem{phonetics_textbook}
J.C. Catford,
\newblock {\em {A Practical Introduction to Phonetics}},
\newblock Oxford University Press, New York, 1988.

\end{thebibliography}

\end{document}